\documentclass[prd,aps,twocolumn,a4paper,showkeys,nofootinbib]{revtex4-1}
\usepackage{amsmath}
\usepackage{amsfonts}
\usepackage{amssymb}	
\usepackage{graphicx}
\usepackage{amsbsy}
\usepackage{color}
\usepackage{commath}
\allowdisplaybreaks
\usepackage{hyperref}
\usepackage{makecell}
\usepackage{multirow}

\def\GMc2{G M_{\odot} c^{-2}}

\def\O{\mathcal{O}}

\def\lm{{\ell m}}

\def\lm{{\ell m}}

\def\lm{{\ell m}}

\def\l{{\ell }}

\def\O{{\cal O}}

\def\ha{{\hat{a}}}

\def\TEOBResumS{\texttt{TEOBResumS}}

\def\Teukode{{\texttt{Teukode}}}

\newcommand\fnp[1]{{\hat{f}_{\varphi #1}^{\rm N _{nc}}}}
\def\NP22{{\fnp{,22}}}

\usepackage{color}
\definecolor{cyan}{rgb}{0,0.9,0.9}
\definecolor{orange}{rgb}{0.9,0.5,0}
\definecolor{magenta}{rgb}{1,0,1}
\definecolor{purple}{rgb}{0.8,0.4,0.8}
\definecolor{gray}{rgb}{0.8242,0.8242,0.8242}
\definecolor{dodgerblue}{rgb}{0.12, 0.56, 1.0}
\definecolor{darkgrey}{rgb}{0.5,0.5,0.5}
\definecolor{darkgreen}{rgb}{0,0.65,0}

\definecolor{colortab1}{rgb}{0.1, 0.1, 1.0}
\definecolor{colortab2}{rgb}{0.9,0,0.1}

\begin{document}

\title{New Avenue for Accurate Analytical Waveforms and Fluxes \\ for Eccentric Compact Binaries}

\author{Simone \surname{Albanesi}${}^{1,2}$}
\author{Andrea \surname{Placidi}${}^{3}$}
\author{Alessandro \surname{Nagar}${}^{2,4}$}
\author{Marta \surname{Orselli}${}^{3,5}$}
\author{Sebastiano \surname{Bernuzzi}${}^{6}$}
\affiliation{${}^{1}$ Dipartimento di Fisica, Universit\`a di Torino, via P. Giuria 1, 
10125 Torino, Italy}
\affiliation{${}^2$INFN Sezione di Torino, Via P. Giuria 1, 10125 Torino, Italy} 
\affiliation{${}^3$Dipartimento di Fisica e Geologia, Universit\`a di Perugia,
INFN Sezione di Perugia, Via A. Pascoli, 06123 Perugia, Italia}
\affiliation{${}^4$Institut des Hautes Etudes Scientifiques, 91440 Bures-sur-Yvette, France}
\affiliation{${}^5$Niels Bohr Institute, Copenhagen University,  Blegdamsvej 17, DK-2100 Copenhagen \O{}, Denmark}
\affiliation{${}^6$Theoretisch-Physikalisches Institut, Friedrich-Schiller-Universit{\"a}t 
Jena, 07743, Jena, Germany}  
\begin{abstract}
We introduce a new paradigm for constructing accurate analytic waveforms (and fluxes) for eccentric compact binaries. 
Our recipe builds on the standard Post-Newtonian (PN) approach but (i) retains implicit time-derivatives of the phase 
space variables in the instantaneous part of the noncircular waveform, and then (ii) suitably factorizes and resums this
partly PN-implicit waveform using effective-one-body (EOB) procedures.
We test our prescription against the exact results obtained by solving the Teukolsky equation with a test-mass source orbiting a 
Kerr black hole, and compare the use of the exact {\it vs} PN equations of motion for the time derivatives computation.  
Focusing only on the quadrupole contribution, we find that the use of the exact equations of motion yields an analytical/numerical 
agreement of the (averaged) angular momentum fluxes that is improved by $40\%$ with respect to previous work, 
with $4.5\%$ fractional difference for eccentricity $e=0.9$ and black hole dimensionless spin $-0.9\leq \hat{a}\leq +0.9$.
We also find a remarkable convergence trend between Newtonian, 1PN and 2PN results. 
Our approach carries over to the comparable mass case using the resummed EOB equations of motion and paves the 
way to faithful EOB-based waveform model for long-inspiral eccentric binaries for current and future gravitational wave detectors.
\end{abstract}
\date{\today}

\maketitle

\section{Introduction}

The accurate computation of the gravitational waves (GW) from eccentric compact binary mergers has recently prompted a vibrant activity by the analytical and numerical relativity community~\cite{Hinder:2017sxy,Hinderer:2017jcs,Chiaramello:2020ehz,Nagar:2020xsk,Islam:2021mha, Nagar:2021gss,Nagar:2021xnh,Albanesi:2021rby,Liu:2021pkr,Yun:2021jnh,Tucker:2021mvo,Setyawati:2021gom,Khalil:2021txt,Ramos-Buades:2021adz,Placidi:2021rkh,Albanesi:2022ywx,Chowdhury:2022hfw,Saini:2022igm}. A main goal of this programme is to provide GW templates for binaries on generic orbits to be used in the next scientific observational runs of the LIGO-Virgo-Kagra (LVK) collaboration~\cite{LIGOScientific:2021djp,LIGOScientific:2022myk,KAGRA:2022fgc} and in the future LISA observations~\cite{LISA:2017pwj,Babak:2017tow,Gair:2017ynp}. Indeed, LVK's GW190521 event indicates that eccentricity might play a fundamental role for the analysis of stellar-mass binary black holes mergers~\cite{LIGOScientific:2020iuh,Gayathri:2020coq,Gamba:2021gap,Romero-Shaw:2021ual}. Similarly, extreme-mass-ratio-inspirals (EMRIs) with
eccentricity are expected to be detected by LISA~\cite{LISA:2017pwj,Babak:2017tow,Gair:2017ynp},

The effective-one-body (EOB) approach to the two body problem~\cite{Buonanno:1998gg,Buonanno:2000ef,Damour:2000we,Damour:2001tu,Damour:2015isa},
and in particular the \TEOBResumS{} model~\cite{Nagar:2020pcj,Riemenschneider:2021ppj,Chiaramello:2020ehz,Nagar:2020xsk,Nagar:2021gss}, 
currently seems the only analytical framework close to delivering a quantitatively accurate model
for arbitrary eccentricity, including dynamical captures. The model currently 
exhibits an acceptable level of flexibility and faithfulness with several exact results obtained 
numerically, from comparable mass binaries to extreme mass ratio inspirals~\cite{Placidi:2021rkh,Albanesi:2022ywx}.
The model is also ready for parameter estimation, as shown for the case of GW190521~\cite{Gamba:2021gap}.
A central issue in the construction of eccentric EOB waveform models  is the accurate 
analytical modelization of the radiation reaction force, i.e. the flux of energy and angular 
momentum emitted by the system that causes the orbit to shrink and circularize.
Reference~\cite{Chiaramello:2020ehz}  introduced the eccentric \TEOBResumS{} by dressing the circular azimuthal component 
of the radiation reaction with the leading-order (Newtonian) noncircular contribution. The same idea is applied to
the waveform~\cite{Chiaramello:2020ehz}.  Within this paradigm, several eccentric EOB models were 
constructed~\cite{Chiaramello:2020ehz,Nagar:2021xnh,Nagar:2021xnh,Nagar:2021gss},
so to support the analysis of GW190521 under the hyperbolic-capture hypothesis~\cite{Gamba:2021gap}. 
We note, however, that its quasi-circular limit is not as accurate as the native quasi-circular 
model~\cite{Nagar:2021xnh,Nagar:2021xnh,Nagar:2021gss}, still delivering unfaithfulnesses $<3\%$ with Numerical Relativity (NR) waveform data.
Building partly upon~\cite{Khalil:2021txt}, the noncircular, factorized and resummed, 
EOB waveform was pushed to 2PN accuracy in~\cite{Placidi:2021rkh}.
\citet{Albanesi:2022ywx} contrasted several proposal for the radiation reaction available in the literature and concluded that 
those based on the Newtonian-factorization (with a 2PN correction) deliver the most faithful representation of the fluxes 
of a test-mass around a Kerr black hole. Reference~\cite{Placidi:2021rkh} used the 2PN-truncated 
equations of motion (EOM) to recast the noncircular correction in
a simpler form, retaining explicit dependence on only the angular and radial momenta $(P_\varphi,P_{R})$. 
Unfortunately, with this approach the behavior at the orbit turning points (apastron and periastron) cannot be 
modified by higher PN corrections, since $P_R=0$ there by definition and all noncircular corrections vanish. 
The accuracy at the radial turning points of the scheme is thus entirely determined  by the Newtonian 
contribution, that explicitly depends on the, there nonvanishing, radial acceleration $\ddot{R}$, see~\cite{Placidi:2021rkh,Albanesi:2022ywx}.

Here we propose a new strategy to significantly improve the behavior of the waveform at the radial 
turning points. The crucial element behind the accuracy of the generic Newtonian prefactor is that 
the derivatives (and notably $\ddot{R}$, see Fig.~17 and 18 of ~\cite{Placidi:2021rkh}) are evaluated using the 
{\it exact} EOM and not the PN-truncated ones.  This is the important keystone that allows one to increase the accuracy 
of Newtonian-like expressions in strong field (basically improving the Ruffini-Wheeler approximation~\cite{Ruffini:1971}, see also~\cite{Nagar:2006xv}).
We extend here this idea to higher PN orders.

\section{From the multipole moments to the factorized waveform}
The multipolar decomposition of the strain waveform reads
$h_+ - i h_\times = {\cal D}_L^{-1}\sum_{\ell=2}^{\infty}\sum_{m=-\ell}^{\ell}h_\lm\; {}_{-2}Y_\lm(\iota,\phi)$,
where ${\cal D}_L$ is the luminosity distance and $ {}_{-2}Y_\lm(\iota,\phi)$ the 
$s=-2$ spin-weighted spherical harmonics.
Following Ref.~\cite{Placidi:2021rkh}, 
the multipolar eccentric waveform is written in factorized form as\footnote{Following standard notation, the parity is $\epsilon=0$ when $\l+m$ is even and $\epsilon=1$ when $\l+m$ is odd.}
$h_\lm = h_\lm^{(N,\epsilon)_{\rm}} \hat{h}_\lm^{\rm c} \hat{h}^{{\rm nc}}_\lm$,
where: 
\begin{enumerate}
	\item[(i)] $h_\lm^{(N,\epsilon)_{\rm}}$ is the generic Newtonian prefactor, 
	first used in Ref.~\cite{Chiaramello:2020ehz} to extend the native quasi-circular \TEOBResumS{} 
	model to the eccentric case.
	\item[(ii)] $\hat{h}_\lm^{\rm c}$ is the PN circular correction, which is then further factorized and resummed 
	following the most recent version of the procedure of Ref.~\cite{Damour:2008gu}, as now implemented in
	the quasi-circular \TEOBResumS{} model~\cite{Nagar:2020pcj,Riemenschneider:2021ppj}.
	\item[(iii)] $\hat{h}^{{\rm nc}}_\lm$ is the  PN noncircular correction, formally introduced in Ref.~\cite{Placidi:2021rkh}. 
	This is further factorized in an hereditary contribution (or tail) and in an instantaneous part,
	$\hat{h}^{{\text{nc}}}_\lm  = \hat{h}^{\text{nc}_{\rm tail}}_\lm \, \hat{h}^{\text{nc}_{\rm inst}}_\lm$.
\end{enumerate}
The computation of the noncircular factors of Ref.~\cite{Placidi:2021rkh} employs as initial input the PN expanded spherical multipoles 
of the waveform of Refs.~\cite{Mishra:2015bqa,Khalil:2021txt}. Depending on the parity of the multipole, these are obtained from the 
corresponding mass-type (when $\ell+m$ is even) or current-type (when $\ell+m$ is odd) radiative spherical multipoles, respectively called 
$U_\lm$ and $V_\lm$,
\footnote{This separation takes place when the motion is planar, namely for spin-aligned 
configurations. In the general case for each multipole one needs both $U_\lm$ and $V_\lm$.} 
which in turn are related to the symmetric trace-free (STF)
source multipole moments by nonlinear PN relations involving time derivatives of the latter. In the case of the $\l=m=2$ mode, 
at 2PN one has\footnote{Throughout the paper we use dots to denote time derivatives.}
$h_{22} = -U_{22}/(\sqrt{2}{}c^4)$, $U_{22} = 2 \sqrt{3} \, \alpha_{22}^{ij} U_{ij}$,  and $U_{ij} =\ddot{ I}_{ij}$,
where $\alpha_{22}^{ij}$ is one of the STF tensors which connect the basis of spherical harmonics to the STF one (see, e.g., Eq.(2.5) of Ref.~\cite{Mishra:2015bqa} for their definition), $U_{ij}$ is the STF radiative mass quadrupole, and $I_{ij}$ is the STF mass quadrupole of the source.
A standard intermediate step in this computation is to systematically replace the time derivatives of the dynamical variables, 
which stem from the time derivatives of the STF source moments, like $\ddot{I}_{ij}$, with the PN-expanded EOM.
Here, we follow a different approach to obtain an alternative noncircular instantaneous factor where the time-derivatives 
are not replaced with the PN-expanded EOM. In practice: 
(i) we recover the 2PN accurate expressions for the STF source moments, valid for noncircular binaries, from Sec.~IIIB of Ref.~\cite{Mishra:2015bqa}; 
(ii) we trade the modified harmonic coordinates used therein for the EOB phase space variables $(r,\varphi,p_{r_*},p_\varphi)$
\footnote{In order of appearance $r\equiv R/M $ is the relative separation in the center of mass frame, $\varphi$ is the orbital phase, $p_{r_*} \equiv (A/B)^{1/2} \, p_r$, where $A,B$ are 
the 2PN accurate EOB potentials and $p_r\equiv P_R/\mu$ is the radial momentum, and  $p_\varphi\equiv P_\varphi / \mu M$ 
is the angular momentum. Denoting with $m_{1,2}$ the individual masses of the binary we have $M \equiv m_1+m_2$, $\mu\equiv m_1 m_2/M$ and $q\equiv m_1/m_2\geq 1$.} 
using the transformations given in Eqs.~(5)–(8) of Ref.~\cite{Placidi:2021rkh}; 
(iii) we build the radiative multipoles following Secs.~II and~IIIA of Ref.~\cite{Mishra:2015bqa}, by taking the needed time 
derivatives but  {\it without} replacing them with the PN-expanded EOM; (iv) we factorize the Newtonian part, which 
is precisely $h_\lm^{(N,\epsilon)_{\rm}}$; (v) we factorize the generic source term $\hat{S}_{\rm eff}^{(\epsilon)}$, which corresponds to the mass-reduced 
EOB effective Hamiltonian $\hat{H}_{\rm eff}$ for even-parity multipoles and to the Newtonian-normalized angular momentum $\hat{j}_\varphi$ 
for odd-parity multipoles; finally (vi) we factorize also the non-eccentric part of the residual, which is obtained by setting to 
zero $p_{r_*}$ and all the time derivatives of the EOB dynamical variable except $\Omega \equiv \dot{\varphi}$. 
Note that this non-eccentric part {\it does not} coincide with the exact circular limit, as we will explain below.
The $\ell=m=2$ PN-expanded instantaneous contribution reads
\begin{figure}[t]
\center
	\includegraphics[width=0.45\textwidth]{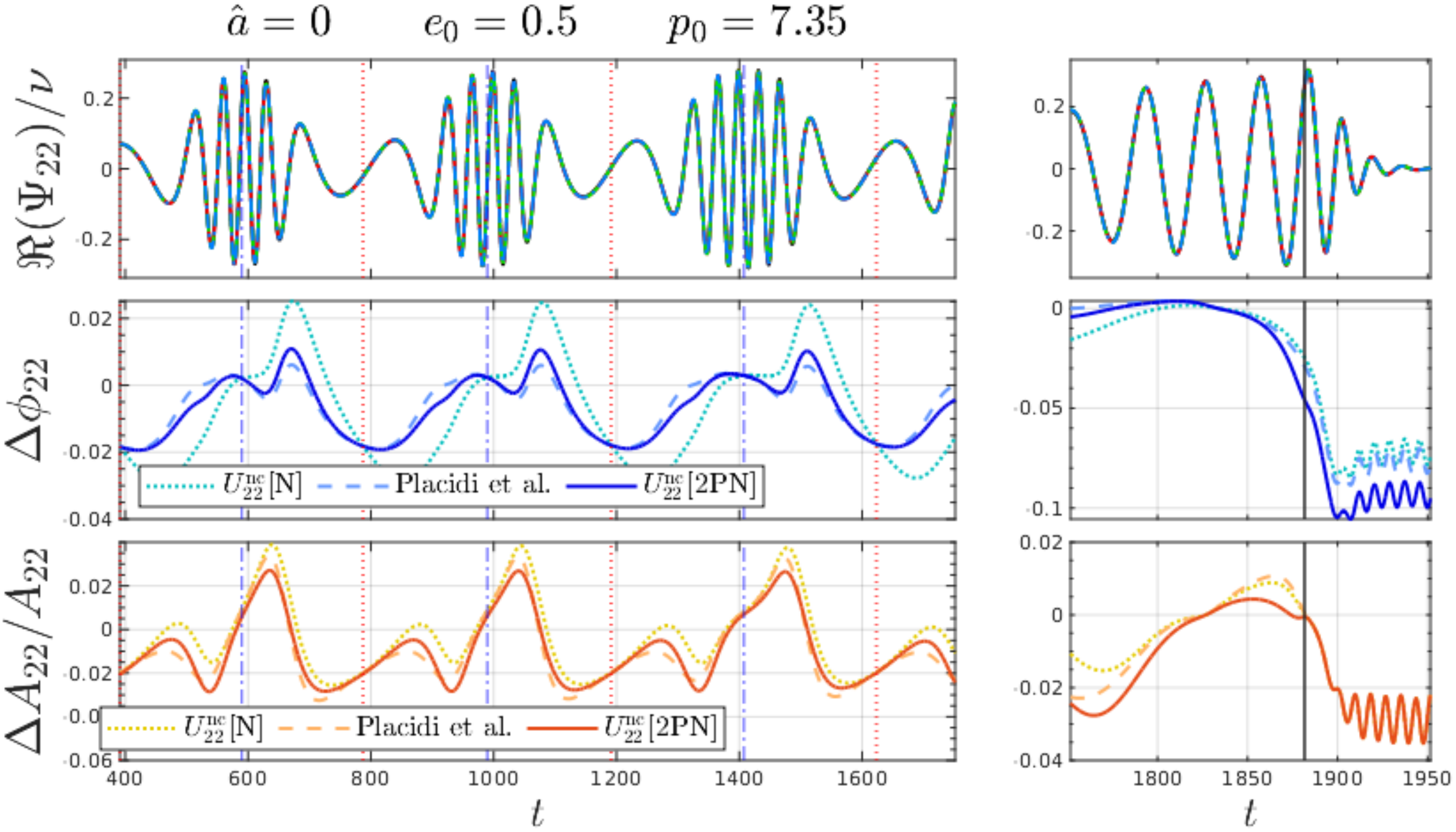}
	\caption{\label{fig:waves} Quadrupolar waveform generated by a test-mass plunging into a Schwarzschild black hole
	along an orbit with initial eccentricity $e_0=0.5$. Numerical waveform (black) compared to the EOB waveform 
	with Newtonian noncircular  corrections (red), to the one with 2PN noncircular corrections~\cite{Placidi:2021rkh} (dashed green), 
	and to the one proposed here (dash-dotted blue). Phase and relative amplitude differences are also shown.
	}
\end{figure}
\begin{align}
\label{eq:22structure}
h^{\text{inst}}_{22} & = h_{22}^{(N,0)}(r,\dot{r},\ddot{r},\Omega,\dot{\Omega}) \\
&+\frac{1}{c^2} h_{22}^{(\rm 1PN,0)}(r,\dot{r},\ddot{r},\Omega,\dot{\Omega},p_{r_*},\dot{p}_{r_*},\ddot{p}_{r_*},p_\varphi,\dot{p}_\varphi,\ddot{p}_\varphi) \cr
&+\frac{1}{c^4} h_{22}^{(\rm 2PN,0)} (r,\dot{r},\ddot{r},\Omega,\dot{\Omega},p_{r_*},\dot{p}_{r_*},\ddot{p}_{r_*},p_\varphi,\dot{p}_\varphi,\ddot{p}_\varphi)\nonumber,
\end{align}
where and $h_{22}^{(N,0)}$ is the Newtonian part and $(h_{22}^{(\rm 1PN,0)},h_{22}^{(\rm 2PN,0)})$ 
formally addresses the contributions obtained by taking the time-derivative of the corresponding terms 
in the radiative multipoles while keeping all derivatives explicit. We obtain
\begin{align}
h_{22}^{(N,0)} & \equiv -8\sqrt{\dfrac{\pi}{5}}\nu r^2 \Omega^2 e^{-2{\rm i}\varphi}
\hat{h}_{22}^{(N, 0)_{\rm nc}}, \\
\hat{h}_{22}^{(N, 0)_{\rm nc}} & =1-\dfrac{1}{2}\left(\dfrac{\dot{r}^2}{r^2\Omega^2} 
+ \dfrac{\ddot{r}}{r\Omega^2}\right) 
+ {\rm i}\left(\dfrac{2\dot{r}}{r\Omega} + \dfrac{\dot{\Omega}}{2\Omega^2}\right), \nonumber
\end{align}
with the noncircular part $\hat{h}_{22}^{(N, 0)_{\rm nc}}$ isolated.
The noncircular contribution is obtained from Eq.~\ref{eq:22structure} as follows.
First, we define the instantaneous, total, factorized correction as
$f_{22}^{\rm total}\equiv h_{22}^{\rm inst}\left(h_{22}^{(N,0)} \hat{H}_{\rm eff}\right)^{-1}$,
where we replaced $\hat{S}^{(0)}_{\rm eff}\equiv \hat{H}_{\rm eff}$.
The  non-eccentric limit of this function is defined according to point (iv) above, so 
to obtain (at 2PN)
\begin{align}
f_{22}^{\rm circ}&= 1+ \dfrac{u}{c^2}\left[-\frac{12}{7}-\frac{p_\varphi^2 u}{3}+\left(\frac{1}{7}+\frac{p_\varphi^2 u}{2}\right) \nu\right]\cr
&+ \dfrac{u^2}{c^4}\left[ -\frac{229}{252}-\frac{929 p_\varphi^2 u}{756}+\frac{19 p_\varphi^4
	u^2}{63}\right. \cr
&\left.+\left(\frac{289}{126}-\frac{1741 p_\varphi^2 u}{378}-\frac{235 p_\varphi^4
	u^2}{504}\right) \nu \right. \cr
&\left.+\left(\frac{65}{126}+\frac{31 p_\varphi^2 u}{54}-\frac{143 p_\varphi^4 u^2}{504}\right) \nu ^2\right].
\end{align}
Note that $p_\varphi$ {\it is not} replaced with its PN-expanded circular expression.
The noncircular (instantaneous) contribution is obtained factoring out this result as
$\hat{h}_{22}^{\rm nc_{inst}} =  T_{\text{2PN}} \left[  f^{\text{\rm total}}_{22}({f}_{22}^{\rm circ})^{-1}\right]$,
where $T_{\text{2PN}}$ indicates an expansion up to the 2PN order. This allows us to obtain a
new noncircular factor that is analogous, though different, to that of Ref.~\cite{Placidi:2021rkh}. 
A few more comments are in order to further clarify the structure of the non-eccentric (i.e. circular) part. First, 
we stress that taking the exact {\it circular limit} of  the product $h^{(N,0)}_{22} f_{22}^{\rm circ}$ 
(i.e., replacing also the PN-expanded expression of $p_\varphi$ along circular orbits) correctly delivers
the 2PN-accurate $f_{22}$ function of Eq.~(B1) of Ref.~\cite{Damour:2008gu}. 
The factorization procedure is thus such that the waveform is consistent with the quasi-circular waveform
of {\TEOBResumS}\footnote{In Ref.~\cite{Placidi:2021rkh} we factorized the {\it 
circular} $\hat{S}_{\rm eff}^{(\epsilon)}$, but the procedure followed in this work is more consistent since
in the full EOB waveform we have the generic factor $\hat{S}_{\rm eff}^{(\epsilon)}$, see e.g. 
Ref.~\cite{Damour:2008gu}.}.
We note, however, that in practice we {\it do not} use the 2PN-accurate $f_{22}^{\rm circ}$ recovered above, 
but rather replace it with the quasi-circular function $\rho_{22}=f_{22}^{1/2}$ with the PN-accuracy and
resummation used either in the standard \TEOBResumS{} model (for comparable masses) 
or in its test-mass version~\cite{Albanesi:2021rby}. More precisely, $\rho_{22}$ is resummed 
according to Refs.~\cite{Nagar:2016ayt, Messina:2018ghh}, but while in \TEOBResumS{} 
the orbital contribution $\rho_{22}^{\rm orb}$ is Taylor-expanded at $3^{+2}$PN 
accuracy~\cite{Damour:2009kr}, in the test-mass limit we 
use it at 6PN, resummed with a (4,2) Pad\'e approximant.

The new noncircular factor is given explicitly in a  supplemental {\tt Mathematica} notebook.
The noncircular tail contribution $\hat{h}^{\text{nc}_{\rm tail}}_\lm$ is resummed following Ref.~\cite{Placidi:2021rkh}, while
the circular tail contributions are incorporated according to standard procedure~\cite{Damour:2008gu}.
From the waveform, the quadrupolar fluxes read~\cite{Nagar:2005ea}
$\dot{E}_{22} = \dfrac{1}{8\pi}|\dot{h}_{22}|^2$ and 
$\dot{J}_{22} =-\dfrac{1}{4\pi} \Im\left(\dot{h}_{22} h_{22}^* \right)$.

\begin{figure}[t]
\center
	\includegraphics[width=0.45\textwidth]{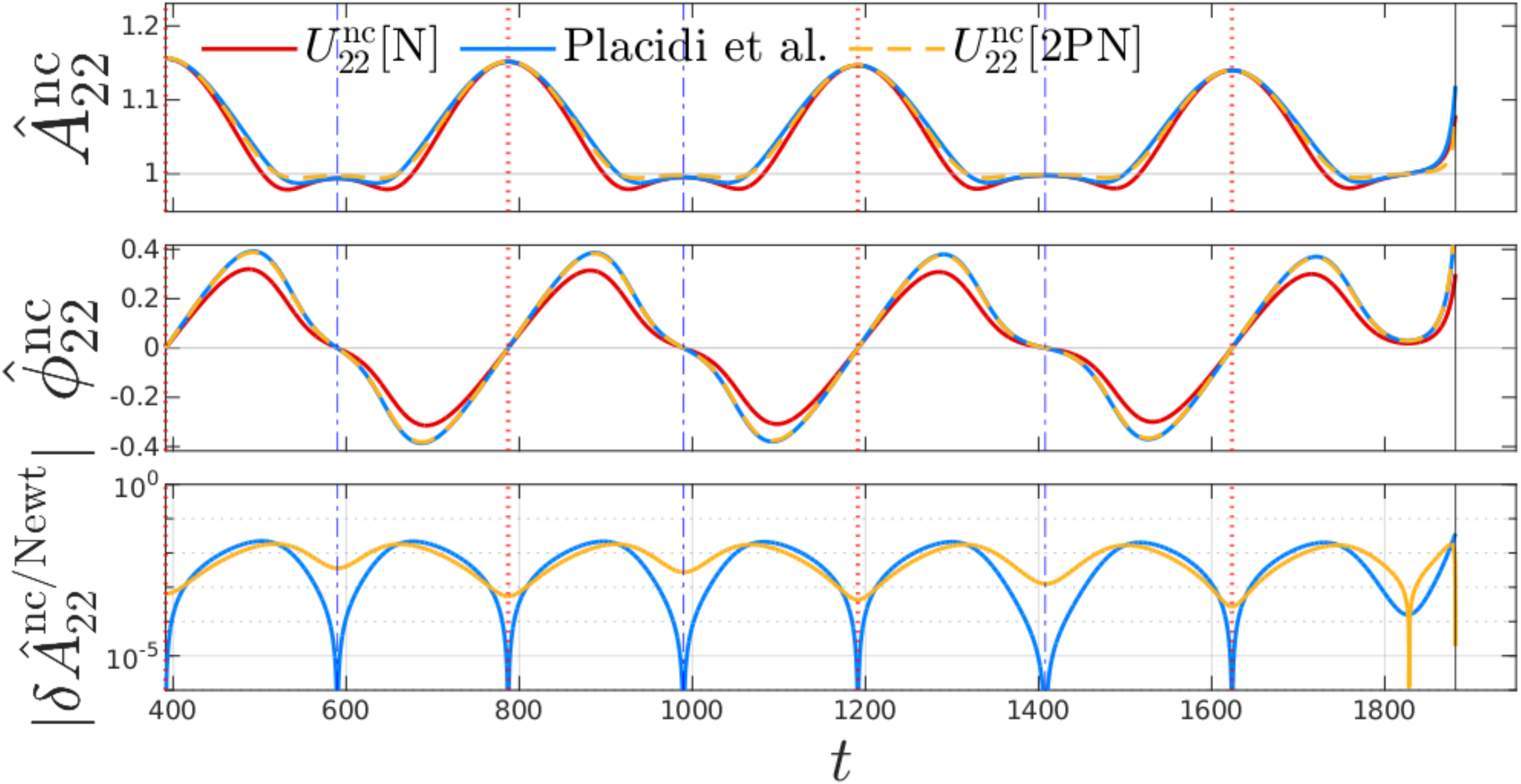}
	\caption{\label{fig:NCfactors} 
	Same configuration of Fig.~\ref{fig:waves}: comparing the 2PN-accurate
	instantaneous noncircular correction to the amplitude (top) and to the phase (middle)
	of this work with the one of Placidi~et al.\cite{Placidi:2021rkh}. Bottom: relative difference with 
	the Newtonian contribution. The noncircular factor  introduced here is nonzero at the apastron (red dotted vertical lines) 
	and periastron (blue dash-dotted vertical lines). 
	The black vertical line marks the merger time.
	}
\end{figure}

\section{Comparing analytical and numerical results}
\label{sec:results}
\begin{figure}[t]
\center
	\includegraphics[width=0.48\textwidth]{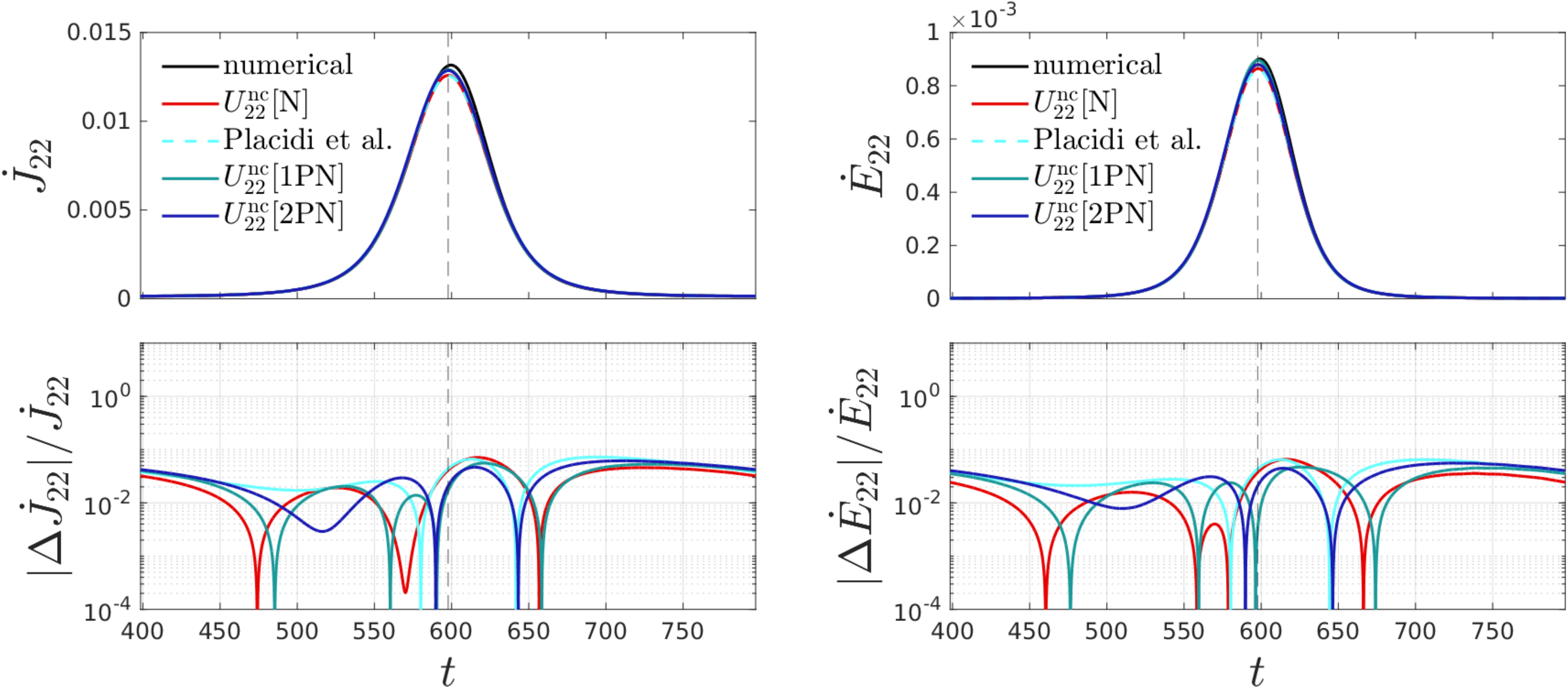}
	\caption{\label{fig:fluxes_e05} Angular momentum and energy 
	quadrupolar fluxes at infinity generated by a test-particle in 
	Schwarzschild spacetime along a geodesic with $e=0.5$ and $p=9$. 
	The relative differences in the bottom panels show that the the PN waveform corrections
	with explicit derivatives improve the analytical/numerical agreement at periastron.}
\end{figure}

The new prescription for the noncircular waveform correction is tested by following
step-by-step the approach of Ref.~\cite{Albanesi:2021rby,Placidi:2021rkh,Albanesi:2022ywx}.
This relies on extensive comparisons with waveform and fluxes emitted by a (nonspinning) 
particle orbiting around a Kerr black hole, considering various orbital configurations.
To set the stage, we consider a particle inspiralling and plunging around a Schwarzschild 
black hole and compare the analytical waveform with the numerical one, considered exact, 
obtained solving numerically the Teukolsky equation using {\Teukode}~\cite{Harms:2014dqa}
(see Ref.~\cite{Albanesi:2021rby} for more numerical details). Figure~\ref{fig:waves} refers to
a configuration with initial eccentricity $e_0=0.5$ and semilatus rectum $p_0=7.35$.
The numerical waveform (black) is compared with the analytical 
waveform with Newtonian noncircular corrections (red), the waveform with 
2PN noncircular corrections proposed in Ref.~\cite{Placidi:2021rkh} (dashed green), and
the waveform with 2PN noncircular corrections as proposed in this work (blue dash-dotted). 
The derivatives of coordinates and momenta which appear in the noncircular corrections
are computed numerically with a $4^{\rm th}$-order centered-stencil scheme.
The corresponding  analytical/numerical differences for amplitude and phase are 
shown in the bottom panels. 
Regarding the phase, the performance of the new waveform and 
of the one of Ref.~\cite{Placidi:2021rkh} are substantially equivalent (see the dashed and
solid blue lines in the middle panel of Fig.~\ref{fig:waves}). 
For the amplitude, instead, the new approach yields a reduced maximum analytical/numerical 
difference during the evolution, as well as a slight improvement as the orbital motion approaches 
the periastra (see bottom panel of Fig.~\ref{fig:waves}). This comes from the nonvanishing of
the noncircular correction at the radial turning point. This is highlighted in Fig.~\ref{fig:NCfactors}, 
which shows the noncircular corrections corresponding to Fig.~\ref{fig:waves}. 
The bottom panel exhibits the relative difference between the two 2PN noncircular 
amplitude corrections and the Newtonian noncircular amplitude correction, 
$|\delta \hat{A}_{22}^{\rm nc/Newt}|$. The correction at the two turning points is
nonzero and is especially relevant at periastron.
Note that the {\it instantaneous} phase noncircular corrections at 2PN differ
sensibly from the Newtonian one. However, part of this difference is compensated
by the hereditary phase correction, as already highlighted in Sec.~IIIC 
of Ref.~\cite{Placidi:2021rkh}.

\begin{figure}[t]
\center
	\vspace{0.1cm}
	\includegraphics[width=0.23\textwidth]{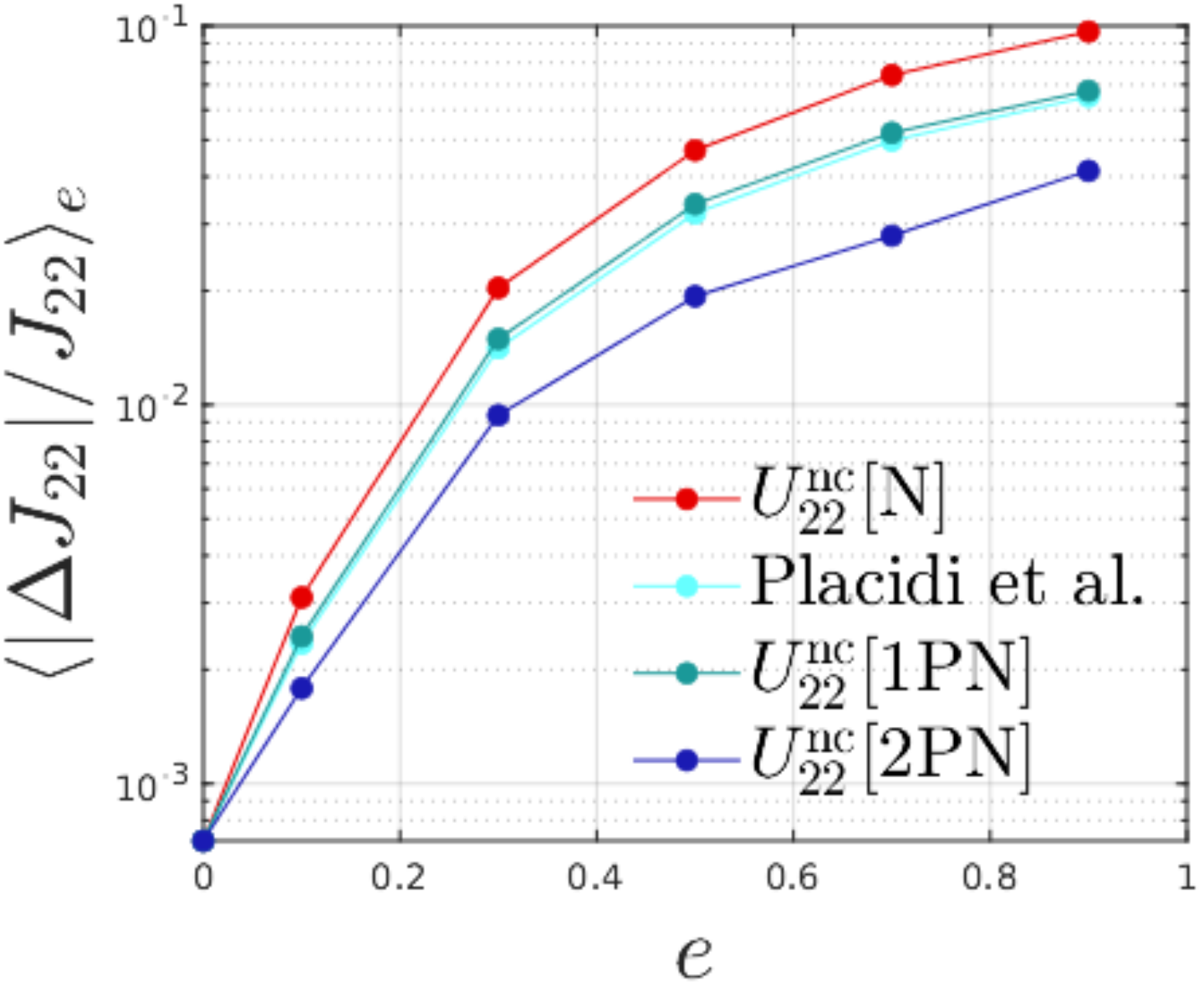}
	\includegraphics[width=0.23\textwidth]{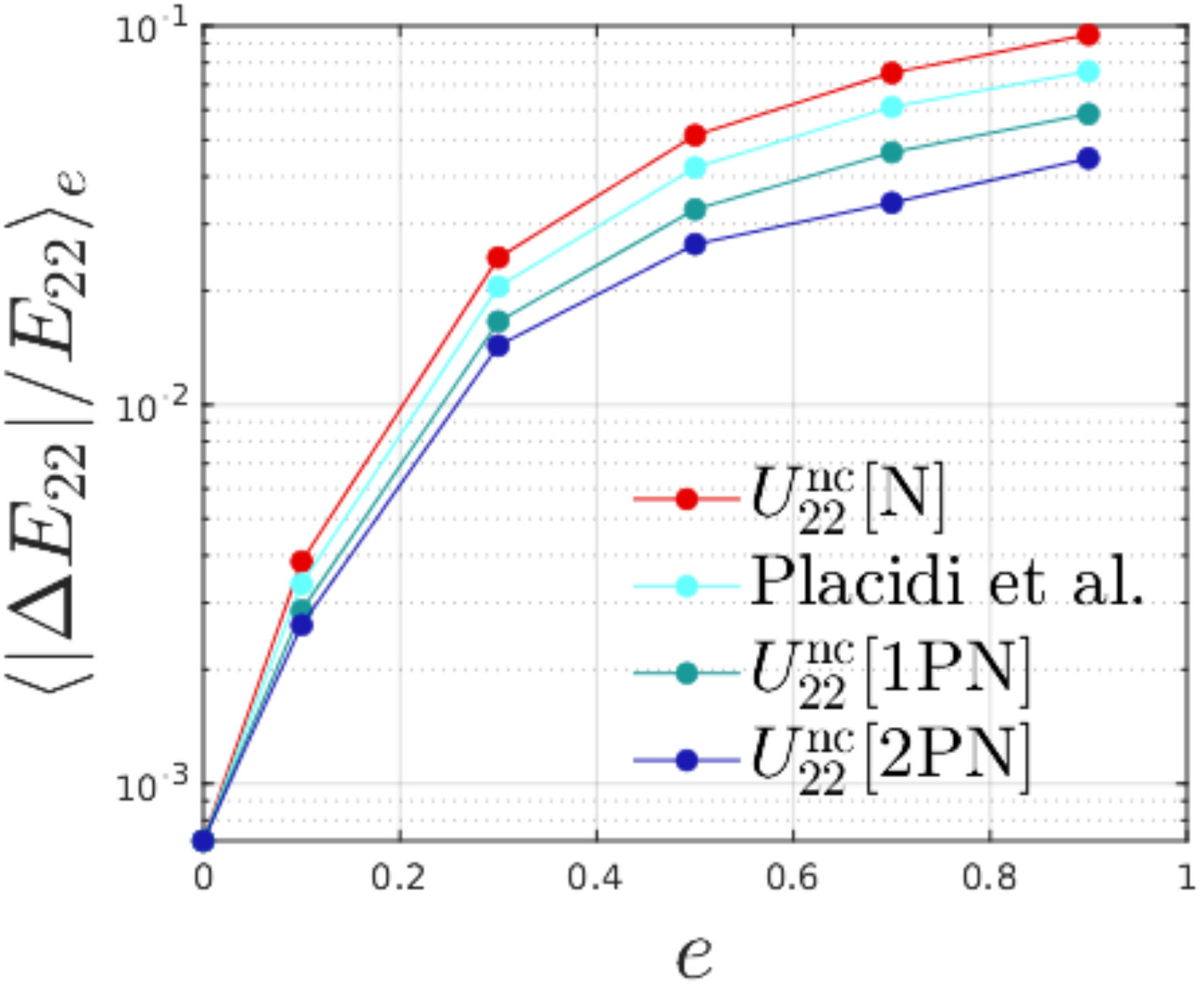}
	\caption{\label{fig:mean_fluxes}
	Analytical/numerical fractional differences between the averaged quadrupolar fluxes 
	versus eccentricity that show convergence of PN corrections in the test-mass limit. 
	Each point is obtained from the mean of the orbital averaged fluxes of 
	all the configurations considered at a given eccentricity $e$.  
	See the main text for more details.
    }
\end{figure}
%
\begin{figure}[t]
\center
	\includegraphics[width=0.45\textwidth]{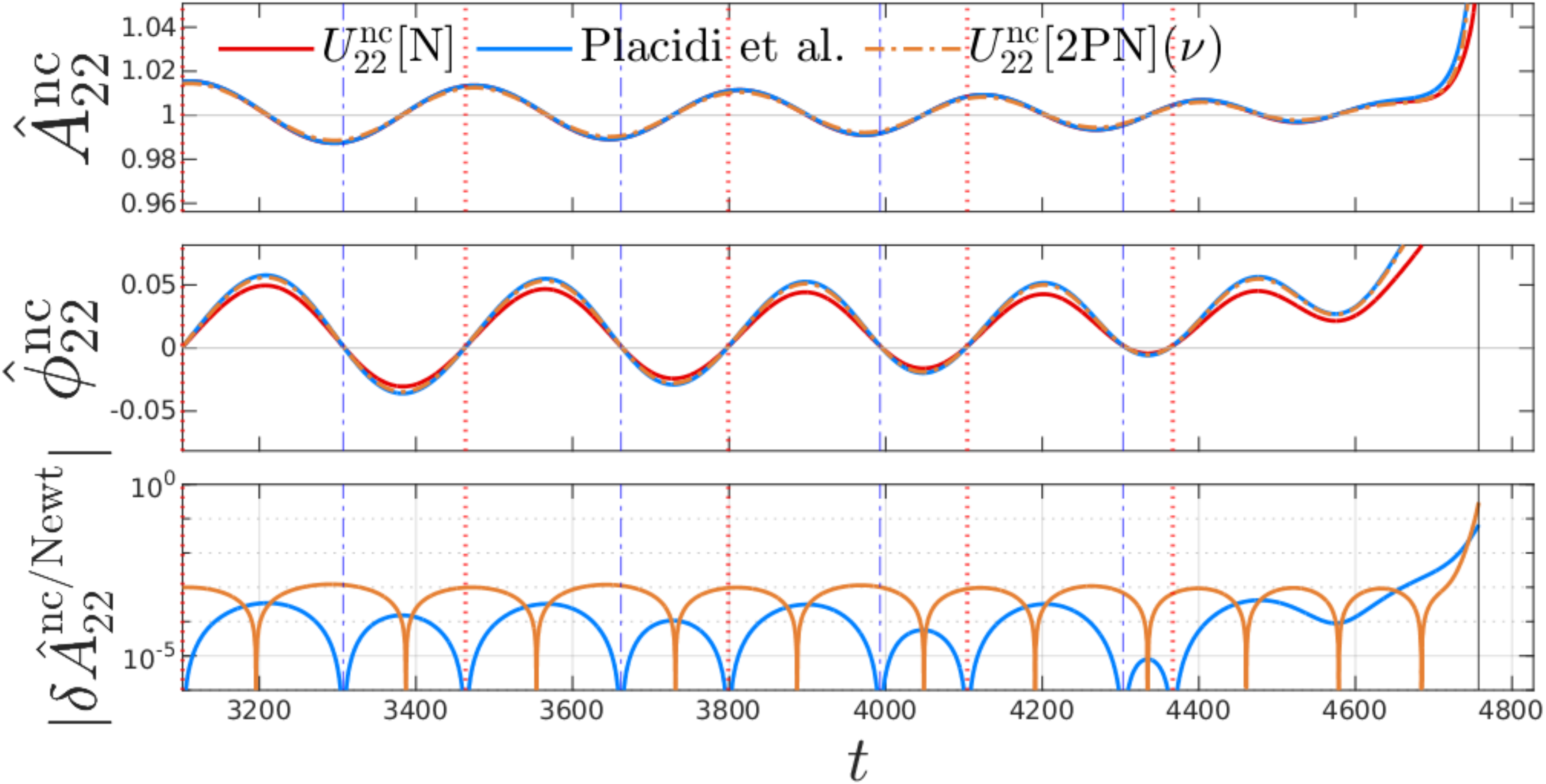}
	\caption{\label{fig:NCfactors_q1d22}
	Same scheme as Fig.~\ref{fig:NCfactors}, but for an eccentric inspiral binary 
	with $e_0=0.07621$ and $q=1.22$ obtained with 
	the full EOB dynamics of ~\cite{Placidi:2021rkh}. 
	}
\end{figure}
The improvement in the description of the waveform at periastron is even more
important when the noncircular corrections are incorporated in the fluxes. Note in
fact that the main contribution to the dynamics, through radiation reaction, happens
due to the burst of radiation emitted at periastron.
To show this effect systematically,  we consider a set of geodesic eccentric orbits 
with dimensionless Kerr spin $|\ha|\leq 0.9$, eccentricity up to $e=0.9$ and 
semilatera recta given by  $p = p_{\rm schw} p_s(e,\ha)/p_s(e,0)$,
where $p_s$ is the separatrix~\cite{OShaughnessy:2002tbu,Stein:2019buj} 
and $p_{\rm schw} = (9, 13)$. The definitions of eccentricity and semilatus rectum used here can 
be found in Ref.~\cite{Albanesi:2021rby}, together with more details on the numerical data. 
In Fig.~\ref{fig:fluxes_e05} we compare the fluxes for a case example with $e=0.5$. 
From the analytical/numerical relative differences one finds that the 2PN noncircular corrections with explicit derivatives
perform better at periastron. As radiation reaction, these
fluxes will drive faster inspirals than those driven by the simple 
leading (Newtonian) noncircular correction of~\cite{Albanesi:2021rby}.
To draw a more global picture, it is useful to compare the orbital-averaged analytical fluxes with 
the corresponding, averaged, numerical ones obtained from the exact waveforms, calculated solving the Teukolsky equation.
We compute the  analytical/numerical relative difference and for each value of eccentricity
$e$ we compute its average over all the dataset sharing the same value of $e$.
These averages are shown in Fig.~\ref {fig:mean_fluxes}, where each point has been obtained by
averaging the analytical/numerical relative differences of 14 simulations with 
$\ha = (0,\pm 0.2, \pm 0.6, \pm 0.9)$ and two different values of $p$.
As shown in Fig.~\ref {fig:mean_fluxes}, the new 2PN  waveform yields (on average) 
the best analytical/numerical agreement; even the 1PN energy flux calculation is
better than the 2PN-accurate expression of Ref.~\cite{Placidi:2021rkh}.
Note however that the average over all spinning configurations can hide some information.
In particular,  for highly eccentric configurations ($e=0.9$), 
the Newtonian prescription yields a better analytical/numerical agreement when averaged 
only on negative spins.
This is clearer for the lowest spin, $\hat{a} = -0.9$, since the Newtonian flux yields a better 
analytical/numerical agreement even for mild eccentricities ($e\gtrsim 0.3$).
However, in the Schwarzschild case, the hierarchy of the different prescriptions 
is the same shown in Fig.~\ref{fig:mean_fluxes}.

The new noncircular correction factor is quantitatively superior to all other previous 
attempts of incorporating high PN noncircular information in the waveform and fluxes. 
This is further corroborated by the following: the instantaneous amplitude correction
presented in Eq.~(37) of Ref.~\cite{Placidi:2021rkh} contains a 1PN term 
$\propto - p_{r_*}^2r$,  that can become extremely large when considering
hyperbolic or eccentric orbits with large radius. While this issue is not relevant for any 
of the configurations considered in Ref.~\cite{Placidi:2021rkh}, 
the correction can become even negative, and thus unphysical,
for large separations, e.g. those occurring in hyperbolic encounters.
By contrast, the new waveform is well-behaved {\it also} for a hyperbolic 
encounter or scattering configuration starting from any, arbitrarily large, initial separation.
\begin{figure}[t]
\center
	\includegraphics[width=0.45\textwidth]{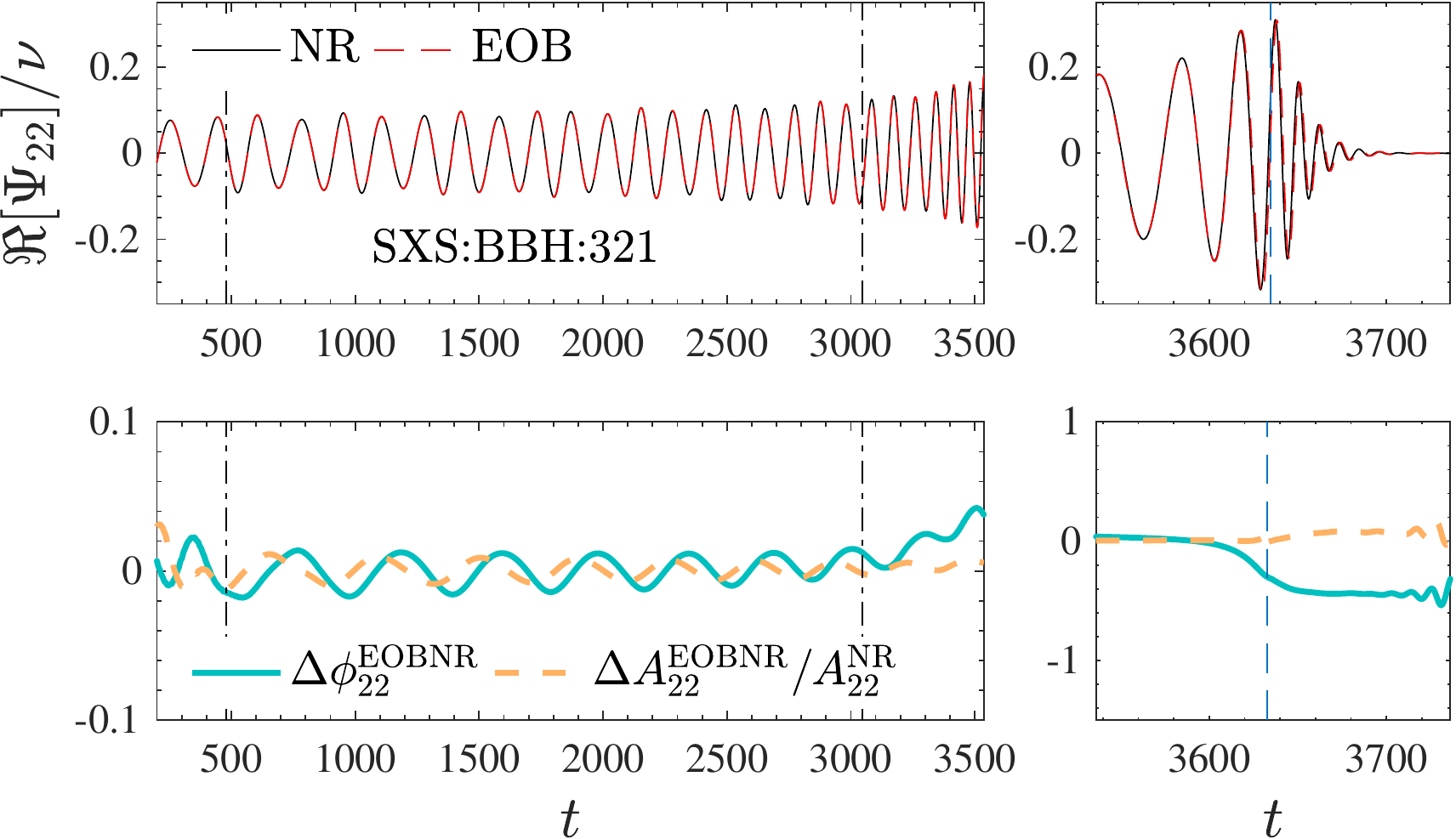}
	\caption{\label{fig:321} Illustrative EOB/NR phasing comparison with the NR dataset SXS:BBH:321
	of the SXS catalog~\cite{SXS:catalog}. The EOB/NR phase difference is reduced during the plunge 
	with respect to the corresponding plot in Fig.~13 of Ref.~\cite{Placidi:2021rkh}. The noncircular waveform 
	correction is shown in Fig.~\ref{fig:NCfactors_q1d22}.
	}
\end{figure}

The same behavior carries over to the comparable-mass case, with the test-mass dynamics 
replaced by the resummed EOB dynamics. Figure~\ref{fig:NCfactors_q1d22} exhibits the time-evolution 
of the noncircular waveform corrections along the EOB dynamics of a binary corresponding to the NR configuration SXS:BBH:321 
of the SXS catalog~\cite{SXS:catalog}, row $\#23$ in Table~IV of Ref.~\cite{Placidi:2021rkh}. In this case,
the mass ratio is $q=1.22$, while the dimensionless spins $(\chi_1,\chi_2)$, aligned with the orbital angular momentum, 
are $\chi_1=+0.33$ and $\chi_2=-0.44$. The initial EOB eccentricity at the apastron is small, 
$e_{\omega_a}^{\rm EOB}=0.07621$, but large enough to probe that our new waveform 
brings an improvement with respect to previous work. First of all, Fig.~\ref{fig:NCfactors_q1d22} 
indicates that in the comparable-mass case the amplitude correction at the radial turning points is 
more relevant than in the test-mass case (cf. Fig.~\ref{fig:NCfactors}), although the correction is small.
It is informative to look at a standard EOB/NR phasing comparison for SXS:BBH:321, that we report
in Fig.~\ref{fig:321}. The EOB waveform is aligned to the NR one by minimizing the phase difference 
in the frequency interval corresponding to the two vertical lines in the left panels of the figure. 
The top panels compare the EOB and NR real parts of the waveform, while the bottom panels show the 
EOB/NR phase difference $\Delta\phi^{\rm EOBNR}_{22}\equiv \phi_{22}^{\rm EOB}-\phi_{22}^{\rm NR}$
and relative amplitude difference, with $\Delta A_{22}^{\rm EOBNR}\equiv A_{22}^{\rm EOB}-A_{22}^{\rm NR}$.
The picture illustrates that $\Delta\phi^{\rm EOBNR}_{22}$ is reduced, during the late-inspiral and plunge,
with respect to the corresponding plot in Fig.~13 of Ref.~\cite{Placidi:2021rkh}, with the same waveform alignement
interval used here. A similar behavior is also found with higher eccentricities. 
However, it must be noted that since the waveform is different, the choice of the initial parameters,
which is not changed in this case, possibly might be optimized further. This study, together with the analysis
of higher modes, is postponed to future work.

We finally point out that the waveform differences due to the new noncircular correction 
will yield fluxes which are {\it larger} at apastron than the current ones.  Once recasted in the form of radiation reaction force, 
and incorporated within the EOB dynamics, the new prescription  will eventually yield an additional acceleration of the eccentric 
inspiral due to the stronger emission at periastron. The development of the radiation reaction force and 
its influence on the inspiral (for any mass ratio)  is also deferred to future work. To do so, we will use, and generalize, 
the approach adopted for the generic Newtonian prefactor~\cite{Chiaramello:2020ehz}. This calculation relies on the 
iterative procedure for computing the time-derivatives including dissipative terms, see Appendix A of Ref.~\cite{Damour:2012ky}.

\section{Conclusion}
We introduced a new way of exploiting PN results for the waveform emitted by 
eccentric and hyperbolic binaries. 
The key idea is to use the full, resummed EOB EOM to compute the time 
derivatives in the formal expression of the analytical waveform, rather than replacing them with 
the PN-expanded EOM. This exploits the robust behavior of the EOB model in the strong-field regime.
This procedure effectively generalizes the use of the generic Newtonian prefactor to higher PN orders. 
We comprehensively tested this approach against a large set of data for waveforms and fluxes emitted 
by a test-mass orbiting a Kerr black hole and we showed that these findings carry over to the comparable
mass case. The novel use of PN results, within the EOB framework, yields considerable 
improvement in the waveform accuracy. Typically, one starts from PN-expanded 
results and then devise specific techniques to improve their behavior in strong field. 
Our findings indicate that this is not sufficient, but it is actually possible to do better by removing some 
of the intermediate steps involving PN-expansions.  In particular, our approach does not rely on 
eccentricity-expanded fluxes, in contrast to~\cite{Munna:2020juq,Munna:2020iju}, and remains
robust also for highly eccentric configurations.
 
 \begin{acknowledgments}
We are indebted to P.~Rettegno for a crucial check on the 2PN waveform of Ref.~\cite{Placidi:2021rkh}
that eventually prompted this study. We are grateful to C.~Munna for informative discussions
and explanations about his PN results.
We also thank Tony Garbato for continuous support during the development of this work.
S.B. acknowledges support by the EU H2020 under ERC Starting Grant, no.~BinGraSp-714626.
M.O. and A.P. acknowledge support from the project ``MOSAICO'' financed by 
Fondo Ricerca di Base 2020 of the University of Perugia. M.O. and A.P. thank the Niels Bohr Institute for hospitality.
Computations were performed on the ``Tullio'' server in Torino, supported by INFN.
\end{acknowledgments}

\bibliography{refs20220831.bib,refs_loc20220831.bib}

\end{document}